\documentclass[prd,preprint,superscriptaddress,nofootinbib,longbibliography,aps]{revtex4-1}
\usepackage[utf8]{inputenc}
\usepackage{graphicx}
\usepackage{mathtools}
\usepackage{longtable}
\usepackage{amsmath}
\usepackage{color} 
\usepackage{amssymb}
\usepackage{subfigure}
\usepackage{microtype}
\usepackage[linktoc=all]{hyperref}
\usepackage{cleveref}
\usepackage{stackengine}
\usepackage{color}
\usepackage{bm}
\usepackage{braket}
\usepackage{ulem,xpatch}
\usepackage{slashed}
\definecolor{myblue}{rgb}{ 0.188, 0.478,0.858}

\newcommand{\be}{\begin{equation}}
\newcommand{\ee}{\end{equation}}

\begin{document}  
\title{Convolutional double copy in (Anti) de Sitter space}

\author{Qiuyue Liang}
\email{qiuyue.liang@ipmu.jp}
\affiliation{Kavli Institute for the Physics and Mathematics of the Universe (WPI), University of Tokyo, Kashiwa 277-8583, Japan}
\affiliation{Center for Particle Cosmology, Department of Physics and Astronomy, University of Pennsylvania, Philadelphia, Pennsylvania 19104, USA}
\author{Silvia Nagy}
\email{silvia.nagy@durham.ac.uk}
 \affiliation{Department of Mathematical Sciences, Durham University, Durham, DH1 3LE, UK}

\date{\today}

\begin{abstract}
The double copy is a remarkable relationship between gauge theory and gravity that has been explored in a number of contexts, most notably scattering amplitudes and classical solutions. The convolutional double copy  provides a straightforward method to bridge the two theories via a precise map for the fields and symmetries at the linearised level. This method has been thoroughly investigated in flat space, offering a comprehensive dictionary both with and without fixing the gauge degrees of freedom. In this paper, we extend this to curved space with an (anti) de Sitter background metric. We work in the temporal gauge, and employ a modified convolution that involves the Mellin transformation in the time direction. As an example, we show that the point-like charge in gauge theory double copies to the (dS-) Schwarzschild black hole solution.
\end{abstract} 

\maketitle

\section{Introduction} 

The double copy is a deep and intensely explored connection between Yang-Mills (YM) and gravity \footnote{See reviews \cite{Adamo:2022dcm,Bern:2019prr} and references within.}. It has achieved its biggest successes in the context of scattering amplitudes \cite{Bern:2010yg,Bern:2010ue,Bern:2008qj,Kawai:1985xq}, which has prompted an expansion in a number of directions including the construction of gravitational solutions from their YM counterparts.

In the perturbative formulation, the majority of results are found in a flat background, where intuition can be imported from the scattering amplitudes program. An important question is how much of the success of the double copy translates to curved backgrounds, which would naturally be of great interest in applications to cosmology, or connections to holography. For some works in this direction, see \cite{Han:2022mze,Alkac:2021bav,Prabhu:2020avf,Bahjat-Abbas:2017htu,Carrillo-Gonzalez:2017iyj,Mei:2023jkb,Armstrong:2020woi,Albayrak:2020fyp,Alday:2021odx,Diwakar:2021juk,Sivaramakrishnan:2021srm,Cheung:2022pdk,Herderschee:2022ntr,Drummond:2022dxd,Farrow:2018yni,Lipstein:2019mpu,Jain:2021qcl,Zhou:2021gnu,Armstrong:2022csc,Lipstein:2023pih}.

The convolutional double copy \cite{Anastasiou:2018rdx,Anastasiou:2014qba,Godazgar:2022gfw,Beneke:2021ilf,Ferrero:2020vww,Borsten:2021zir,Borsten:2020xbt,Luna:2020adi,Borsten:2019prq,LopesCardoso:2018xes,Cardoso:2016amd,Cardoso:2016ngt} can be seen most simply as a consequence of the fact that the double copy is naturally a product in momentum space, which becomes a convolution in coordinate space upon taking a Fourier transformation \footnote{Certain families of classical solutions have been formulated as double copies via a position space product (a programme which usually goes under the name of classical double copy \cite{Monteiro:2014cda,Luna:2018dpt}). This is now understood to be related to the additional symmetries present in these systems \cite{Monteiro:2020plf,Monteiro:2021ztt,Luna:2022dxo}.}. Though only formulated at the level of linearised fluctuations, it has the advantage of being a very direct map, applicable for general, arbitrary fluctuations of YM and gravity, when working in the Becchi-Rouet-Stora-Tyutin (BRST) formulation. It is also compatible with the symmetries and dynamics of both theories, and allows for control in gauge choices. It has been explored in flat backgrounds with some preliminary extensions to  homogeneous manifolds \cite{Borsten:2021zir,Borsten:2019prq}. With this tool in hand, we want to explore the extension of the double copy to other background spacetimes. In this paper, we will specifically focus on (A)dS backgrounds, of obvious interest to cosmology/holography.

To achieve this, we need a convolution in (A)dS space. Recall that the convolution on a flat background
\be
[f\star g](x)=\int d^4y f(0+x) g(y-x)\ ,
\label{fstarg}
\ee
relies heavily on both spatial and temporal translational symmetry, which acts transitively on flat space. It is well known that there is no time translational symmetry in de Sitter space, therefore, we will have to modify the convolution to capture the scaling symmetry associated with the dilatation subgroup of the (anti) de Sitter group.     

As pointed out in
\cite{Sleight:2019hfp,Sleight:2019mgd,Sleight:2020obc,Sleight:2021plv,Sleight:2021iix}, following the discussion of AdS/CFT correspondence \cite{Maldacena:1997re,Gubser:1998bc,Witten:1998qj} and the Mellin space representation developed in the context of  CFT \cite{Mack:2009mi,Penedones:2010ue,Paulos:2011ie,Giombi:2017hpr,Sleight:2018epi}, the scaling symmetry of (A)dS space indicates that the proper basis to analyze the mode functions in the time direction is in Mellin space, 
\begin{eqnarray}
   h(\eta)=\int \frac{d s}{2 \pi i} 2 \tilde{h}(s) \eta^{-2 s+\frac{d}{2}}\ , \quad \tilde{h}(s)=\int_0^{\infty} d \eta h(\eta) \eta^{2 s-\frac{d}{2}-1}\ . 
\end{eqnarray}
We therefore define the {\it convolution} along the time direction through 
\begin{eqnarray}
\label{eq,convoMellin}
    h(\eta) \equiv [f \star g] (\eta)=\int \frac{d \tau}{\tau} f(1\times \tau) g\left(\frac{\eta}{\tau}\right)\ .
\end{eqnarray}
One can verify that the convolution $h(\eta)$ in Mellin space is the product of $\tilde f(s)$ and $\tilde g(s)$, 
\begin{eqnarray}
    \tilde{h}(s)=\tilde{f}(s) \tilde{g}(s) \ .
\end{eqnarray}
Therefore, in (A)dS space, one should apply the Mellin transformation to the time direction, and the normal $3D$ Fourier transformation to the spatial directions. 

This requires a separation between the temporal components and the spatial components, and we will see that the {\it temporal gauge} is best suited for this task. This is a non-covariant gauge choice that has been well explored in the literature (see \cite{RevModPhys.59.1067} for a nice review), and was given several different names historically, including Weyl gauge\cite{weyl1950theory}, Heisenberg-Pauli gauge\cite{Pervushin:1993tk}, and synchronous gauge \cite{Ma:1995ey}. The last name is best known for the study of cosmological perturbations and gravitational waves in the expanding universe ever since Lifshitz in 1946 \cite{Lifshitz:1945du}. We will show later in Sec \eqref{sec:ds space} that this is the correct  gauge for us to build the convolutional double copy in de Sitter space.

 The paper is organised as follows: we will start in Sec.\eqref{sec:review} with a brief review of the convolutional double copy in flat space. We extend earlier work to incorporate a general gauge fixing term in the action. We then give the explicit  double copy dictionary with two gauge choices, the Lorenz gauge and the temporal gauge. We continue by generalising the discussion to curved space, mainly focusing on de Sitter space, in Sec.\eqref{sec:ds space}. Here we discuss the implementation of the temporal gauge, and the convolutional double copy dictionary in de Sitter space, as well as the extension to anti de Sitter. In Sec.\eqref{sec:solution}, we use some physical solutions as examples to illustrate our method. We recover the double copy between the (dS-) Schwarzschild black hole solution and the point charge solution in gauge theory in temporal gauge. Throughout the paper, we use the mostly plus signature.

\section{BRST convolution in flat space with a general gauge-fixing functional}
\label{sec:review}
We begin with a review of the main properties of the flat space convolutional double copy.  This has been previously constructed in the BRST context with a Lorenz gauge fixing functional. Here we will extend this to more general gauge choices, and in particular we will allow ourselves to break Lorentz covariance.

A crucial aspect of the convolutional product is that it does not respect the Leibniz rule:
\begin{equation}
\partial_\mu(f\star g)=(\partial_\mu f)\star g=f\star (\partial_\mu g)\ .
\label{derivrule}
\end{equation}

This will be important in deriving the gravitational symmetries via the double copy, and will be another crucial consideration  in the extension to curved backgrounds. 

We also introduce the $\circ$ product, defined as
\begin{equation}
A_\mu\circ A_\nu\equiv A_\mu^a\star \left(\Phi^{-1}\right)^{aa'}\star A_\nu^{a'}\ ,
\label{circprod}
\end{equation}
where we have taken the convolution inverse of the biadjoint scalar field $\Phi^{aa'}$. This is the appropriate building block for the double copy, as it connects with the amplitudes formulation, as further detailed in \cite{Anastasiou:2018rdx,Borsten:2019prq,Luna:2020adi}. It also gives the correct mass dimension, upon normalising all physical fields to have dimension 1. This is easiest seen by going to momentum space, where the convolution becomes a product.   

We will be working in the BRST formalism, with the YM BRST complex
$(A_\mu,c,\bar{c})$, where $c$ and $\bar{c}$ denote the ghost and anti-ghost, respectively. The Lagrangian, upon integrating out the Lautrup-Nakanishi Lagrange multiplier field, can be written as 
\begin{equation} \label{gen_BRST_YM_Lag}
\mathcal{L}=\text{Tr}\left(-\frac{1}{4}F^{\mu\nu}F_{\mu\nu}+\frac{1}{2\xi}\left(n^\mu A_\mu\right)^2 -\bar{c}\, n^\mu\partial_\mu c
 - j_\mu A^\mu - \bar c j_{\bar c} - c j_c
\right) \ . 
\end{equation}
Note that we have been agnostic in our choice of the BRST gauge-fixing functional:
\be 
G[A_\mu]=n^\mu A_\mu\ ,
\ee 
where $n^\mu$ denotes either a differential operator or a (Lorentz-breaking) constant vector. For the Lorenz gauge fixing functional, we will have $n^\mu=\partial^\mu$, whereas the axial gauge or temporal gauge will be denoted by a constant vector $n^\mu$. We also note that the Lagrangian is a slightly modified version of the standard BRST one, as it includes sources for the ghosts. These were shown to be important in disentangling the degrees of freedom of the gravitational fields resulting from the double copy \cite{Anastasiou:2018rdx,Borsten:2019prq,Luna:2020adi}.

The Lagrangian \eqref{gen_BRST_YM_Lag} is invariant under the BRST transformations 
\begin{equation}
\mathcal{Q}A_\mu=\partial_\mu c,\quad \mathcal{Q}c=0,\quad 
\mathcal{Q}\bar{c}=\frac{1}{\xi}n^\mu A_\mu\ ,
\label{YMBRST}
\end{equation}
and the goal is to write a dictionary for the gravitational fields, built out of terms of the type given in \eqref{circprod}, which can reproduce the appropriate gravitational BRST symmetries upon using \eqref{YMBRST}. The gravity BRST system will comprise the graviton $h_{\mu\nu}$, together with its ghost $c_\mu$ and anti-ghost $\bar{c}_\mu$, as well as the dilaton. The two-form sector can be avoided by choosing the two YM complexes entering the double copy to be identical\footnote{Up to a proportionality constant.}.  

We propose the following dictionary in the presence of a general BRST gauge-fixing functional \footnote{We have assumed the norm of $n^\mu$ is non-zero. A different approach would be needed for e.g. light-cone gauge, where $n^\mu$ is null.}: 
\be \label{dict_gen_gauge}
\begin{aligned}
h_{\mu\nu}=&\mathfrak{a_1} A_{\mu}\circ A_{\nu} + \mathfrak{a_2}\frac{1}{n\partial} \partial_{(\mu}A_{\nu)}\circ nA
+ \mathfrak{a_3}\frac{\partial_\mu\partial_\nu}{(n\partial)^2}nA\circ nA 
+ \mathfrak{a_4}\frac{n_{(\mu}\partial_{\nu)}}{n^2n\partial} nA\circ nA\\
&+\mathfrak{a_5}\frac{\partial_\mu\partial_\nu}{\square}A^\rho\circ A_\rho
+\mathfrak{a_6}\xi\frac{\partial_\mu\partial_\nu}{n\partial}c\circ\bar{c}
+ \mathfrak{a_7} \xi\frac{n_{(\mu}\partial_{\nu)}}{n^2}c\circ\bar{c}\\
&+\eta_{\mu\nu}\left(\mathfrak{a}_8\left[A^\rho\circ A_\rho-\frac{1}{\square}\partial A\circ \partial A\right]+\mathfrak{a}_9\frac{1}{n^2}\left[nA\circ nA+2\xi\, 
n\partial c\circ \bar{c}\right]\right)\ ,\\
\varphi=&\mathfrak{b}_1\left[A^\rho\circ A_\rho-\frac{1}{\square}\partial A\circ \partial A\right]+\mathfrak{b}_2\frac{1}{n^2}\left[nA\circ nA+2\xi\, 
n\partial c\circ \bar{c}\right] \ ,
\end{aligned}
\ee 
where we used the shorthands $n\partial\equiv n^\rho\partial_\rho$, $nA\equiv n^\rho A_\rho$ and $n^2\equiv n_\rho n^\rho$. 
The above is chosen such that all the arbitrary numerical parameters $(\mathfrak{a}_1,...)$ are dimensionless\footnote{Allowing the gauge-fixing operator $n^\mu$ to have arbitrary mass dimension $d_n$, we have 
\be 
[A_\mu]=1\ ,\quad [c]=0\ ,\quad [\bar{c}]=3-d_n\ ,\quad [\xi]=2d_n-2\ .
\ee
}.
The virtue of \eqref{dict_gen_gauge} is that it reproduces 
\begin{equation}
\mathcal{Q}h_{\mu\nu}=\partial_{\mu}c_{\nu}+\partial_\nu c_\mu,\quad
\mathcal{Q}\varphi=0\ ,
\label{gravBRST}
\end{equation}
as needed, upon applying \eqref{YMBRST} to the factors. It also allows us to read off the gravity ghost dictionary
\be 
c_\mu=\left(\mathfrak{a}_1+\tfrac{1}{2}\mathfrak{a}_2\right) c\circ A_\mu
+\left(\tfrac{1}{2}\mathfrak{a}_2+\mathfrak{a}_3-\tfrac{1}{2}\mathfrak{a}_6 \right)\frac{\partial_\mu}{n\partial}c\circ nA
+\left(\mathfrak{a}_4-\tfrac{1}{2}\mathfrak{a}_7\right)\frac{n_\mu}{n^2}c\circ nA +\mathfrak{a}_5\frac{\partial_\mu}{\square}c\circ\partial A\ ,
\ee 
satisfying, as required
\be 
\mathcal{Q} c_\mu=0\ ,
\ee 
where we assumed $c\circ n^\rho A_\rho=n^\rho \left(c\circ A_\rho\right)$, which is satisfied both when $n_\mu$ is a differential operator or a constant vector. The anti-ghost is then obtained by conjugating the factors: 
\be \label{antighost_gen}
\bar{c}_\mu=\left(\mathfrak{a}_1+\tfrac{1}{2}\mathfrak{a}_2\right) \bar{c}\circ A_\mu
+\left(\tfrac{1}{2}\mathfrak{a}_2+\mathfrak{a}_3-\tfrac{1}{2}\mathfrak{a}_6 \right)\frac{\partial_\mu}{n\partial}\bar{c}\circ nA
+\left(\mathfrak{a}_4-\tfrac{1}{2}\mathfrak{a}_7\right)\frac{n_\mu}{n^2}\bar{c}\circ nA +\mathfrak{a}_5\frac{\partial_\mu}{\square}\bar{c}\circ\partial A\ .
\ee 

At this point we pause to explain the difference between the BRST notion of gauge fixing, and the classical notion encountered for example in the study of classical solutions. Note that the \textit{BRST gauge fixing} consists of adding a gauge-symmetry breaking term to the Lagrangian (see \eqref{gen_BRST_YM_Lag}), without imposing a condition on the gauge field. One of the upshots of the convolutional dictionary above is that it allows us to derive the gravitational BRST gauge fixing from the YM one \cite{Borsten:2019prq,Luna:2020adi}. This is achieved by recalling the transformation rule
\begin{equation}
\mathcal{Q}\bar{c}_\mu=\frac{1}{\xi}G_\mu[h_{\mu\nu},\varphi]\  ,  
\end{equation}
which allows us to read off the gravitational BRST gauge-fixing functional by direct computation from \eqref{antighost_gen}, upon making use of the dictionaries in \eqref{dict_gen_gauge}.

Additionally, when studying classical solutions, we may wish to impose that \textit{physical gauge fixing} is preserved by the convolutional double copy. This means that we require plugging YM fields satisfying
\be 
G[A_\mu]=n^\mu A_\mu =0
\ee 
results in a gravity field satisfying a desired gauge choice
\be 
G_\mu[h_{\mu\nu},\varphi]=0\ .
\ee 

This will typically impose some constraints on the parameters $(\mathfrak{a}_1,...)$ appearing in the dictionary Eq.\eqref{dict_gen_gauge}. The procedure was described for YM Lorenz and gravity de Donder gauge, respectively, in \cite{Luna:2020adi}. We will briefly summarise the results in \autoref{Example I: Lorenz gauge}. Then in \autoref{flat_axial_subsect} we will extend this to the temporal gauge\cite{Leibbrandt:1987gp,Palumbo:1990kg,Landshoff:1993da,Leibbrandt:1994np,Joglekar:1999zq,Joglekar:1999zt,Scheihing-Hitschfeld:2022xqx}, where $n^\mu=(1,0,0,0)$, and thus the temporal component of the gauge field is set to zero. This is the first example of adopting non-covariant gauge choices \cite{Leibbrandt:1987qv,Leibbrandt1994NoncovariantGQ} in the convolutional double copy dictionary, and will help us to generalize to curved spacetimes.

\subsection{Example I: Lorenz gauge $(n^\mu=\partial^\mu)$} \label{Example I: Lorenz gauge}
We give a brief review of the convolution dictionary in Lorenz gauge for YM, which corresponds to setting $n^\mu = \partial^\mu$. This has been extensively studied in previous literature\cite{Anastasiou:2018rdx,Anastasiou:2014qba,Godazgar:2022gfw,Beneke:2021ilf,Ferrero:2020vww,Borsten:2021zir,Borsten:2020xbt,Luna:2020adi,Borsten:2019prq,LopesCardoso:2018xes,Cardoso:2016amd,Cardoso:2016ngt}, with the physical gauge fixing specifically discussed in \cite{Luna:2020adi}.

Setting $n^\mu=\partial^\mu$ reduces the dictionary in \eqref{dict_gen_gauge} to 
\be \label{dict_Lorenz_gauge}
\begin{aligned}
h_{\mu\nu}=&\mathfrak{a_1} A_{\mu}\circ A_{\nu} + \mathfrak{a_2}\frac{1}{\square} \partial_{(\mu}A_{\nu)}\circ \partial A
+ \left(\mathfrak{a_3}+\mathfrak{a_4}\right)\frac{\partial_\mu\partial_\nu}{\square^2}\partial A\circ \partial A 
+\mathfrak{a_5}\frac{\partial_\mu\partial_\nu}{\square}A^\rho\circ A_\rho\\
&+\left(\mathfrak{a_6}+\mathfrak{a_7}\right)\xi\frac{\partial_\mu\partial_\nu}{\square}c\circ\bar{c}
+\eta_{\mu\nu}\left(\mathfrak{a}_8 A^\rho\circ A_\rho
+\left(\mathfrak{a}_9-\mathfrak{a}_8\right) \frac{1}{\square}\partial A\circ \partial A 
+2\mathfrak{a}_9\xi\, c\circ \bar{c}\right)\ ,\\
\varphi=&\mathfrak{b}_1 A^\rho\circ A_\rho+\left(\mathfrak{b}_2-\mathfrak{b}_1\right)\frac{1}{\square}\partial A\circ \partial A+\mathfrak{b}_2 2\xi\, 
 c\circ \bar{c}\ .
\end{aligned}
\ee 

These indeed match the most general covariant dictionary, as given in \cite{Luna:2020adi}, upon some relabeling of the numerical parameters. As shown in \cite{Luna:2020adi}, requiring that Lorenz gauge for YM leads to de Donder gauge for gravity (both at the level of BRST symmetry and in the physical sense), leads to a constraint on a subset of the parameters in Eq.\eqref{dict_Lorenz_gauge}. The remaining free parameters can then be chosen to give a simple dictionary
\begin{equation}\label{simple_lorenz}
\bar{h}_{\mu\nu}=2A_\mu\circ A_\nu-\frac{2}{\Box}\partial_{(\mu}A_{\nu)}
\circ \partial A\ ,\quad \varphi=2A^\rho\circ A_\rho +4\xi c\circ\bar{c}\ ,
\end{equation}
where we used the notation $\bar{h}_{\mu\nu}=h_{\mu\nu}-\tfrac{1}{2}\eta_{\mu\nu}h^\rho_{\ \rho}$, so that the de Donder gauge condition is given by $\partial^\mu\bar{h}_{\mu\nu}=0$. 
 
Then it is clear that the dictionary in \eqref{simple_lorenz} satisfies
\be 
\partial A=0\quad \Rightarrow \quad \partial^\mu\bar{h}_{\mu\nu}=0\ .
\ee 
As an example, it was shown in \cite{Luna:2020adi} that the point charge in pure Yang-Mills theory,
\begin{equation}
A^a_\mu=\frac{g \alpha^a}{4\pi r}u_\mu\ ,\quad u_\mu=(1,0,0,0)\ ,
\label{Amuform}
\end{equation}
where $\alpha^a$ is a normalised constant color vector, together with the choice of ghosts:
\begin{equation}
c^a=\frac{gD \alpha^a}{4\pi r},\quad \bar{c}^a=\frac{g\bar{D}\alpha^a}{4\pi r},
\label{ghostsol}
\end{equation}
where $D$ and $\bar{D}$ are constant Grassmann numbers, and spectator field
\be 
\Phi^{aa'}=\frac{g\delta^{aa'}}{4\pi r}\ ,
\ee
can reproduce the JNW solution:
\be 
\bar{h}=\frac{\kappa}{2}\frac{M}{4\pi r}u_\mu u_\nu\ ,\quad 
\varphi=-\frac{\kappa}{2}\frac{Y}{4\pi r} \ ,
\ee 
with $M$ and $Y$ being constant parameters. In particular, it is possible to choose $D$ and $\bar{D}$ such that $Y$ vanishes, so that we can recover the Schwarzschild solution in de Donder gauge.

\subsection{Temporal gauge $(n^\mu=(1,0,0,0))$} 
\label{flat_axial_subsect}
Temporal gauge is a special case of non-covariant gauges, with $n^\mu=(1,0,0,0)$ being a constant vector. We will see later in \autoref{sec:ds space} that it allows us to generalize to de Sitter space. We will start with a discussion of temporal gauge in flat space as a warm-up. 

Inserting $n^\mu=(1,0,0,0)$, the BRST Lagrangian in Eq. \eqref{gen_BRST_YM_Lag} now becomes
\begin{equation} 
\mathcal{L}=\text{Tr}\left(-\frac{1}{4}F^{\mu\nu}F_{\mu\nu}+\frac{1}{2\xi}\left(A_0\right)^2 -\bar{c}\, \partial_0 c
  - j_\mu A^\mu - \bar c j_{\bar c} - c j_c
 \right)\ .
 \end{equation}
This is invariant under the BRST transformations 
\begin{equation}
     \mathcal{Q} A_\mu = \partial_\mu c\ ,\ \mathcal{Q} c = 0\ , \ \mathcal{Q} \bar c
     = \frac{1}{\xi} n A 
     = \frac{1}{\xi} A_0 \ .
 \end{equation}
The dictionary in \eqref{dict_gen_gauge} becomes
\be \label{dict_axial_gauge}
\begin{aligned}
h_{\mu\nu}=&\mathfrak{a_1} A_{\mu}\circ A_{\nu} + \mathfrak{a_2}\frac{1}{\partial_0} \partial_{(\mu}A_{\nu)}\circ A_0
+ \mathfrak{a_3}\frac{\partial_\mu\partial_\nu}{(\partial_0)^2}A_0\circ A_0 
- \mathfrak{a_4}\frac{n_{(\mu}\partial_{\nu)}}{\partial_0} A_0\circ A_0\\
&+\mathfrak{a_5}\frac{\partial_\mu\partial_\nu}{\square}A^\rho\circ A_\rho
+\mathfrak{a_6}\xi\frac{\partial_\mu\partial_\nu}{\partial_0}c\circ\bar{c}
- \mathfrak{a_7} \xi n_{(\mu}\partial_{\nu)}c\circ\bar{c}\\
&+\eta_{\mu\nu}\left(\mathfrak{a}_8\left[A^\rho\circ A_\rho-\frac{1}{\square}\partial A\circ \partial A\right]-\mathfrak{a}_9\left[A_0\circ A_0+2\xi\,
\partial_0 c\circ \bar{c}\right]\right)\ ,\\
\varphi=&\mathfrak{b}_1\left[A^\rho\circ A_\rho-\frac{1}{\square}\partial A\circ \partial A\right]-\mathfrak{b}_2\left[A_0\circ A_0+2\xi\, 
\partial_0 c\circ \bar{c}\right]\ ,
\end{aligned}
\ee 
where $n_\mu=(-1,0,0,0)$. This again reproduces the appropriate transformations \eqref{gravBRST} and the associated graviton ghost takes the form
\be 
c_\mu=\left(\mathfrak{a}_1+\tfrac{1}{2}\mathfrak{a}_2\right) c\circ A_\mu
+\left(\tfrac{1}{2}\mathfrak{a}_2+\mathfrak{a}_3-\tfrac{1}{2}\mathfrak{a}_6 \right)\frac{\partial_\mu}{\partial_0}c\circ A_0
-\left(\mathfrak{a}_4-\tfrac{1}{2}\mathfrak{a}_7\right)n_\mu c\circ A_0 +\mathfrak{a}_5\frac{\partial_\mu}{\square}c\circ\partial A\ ,
\ee 
which satisfies $\mathcal{Q} c_\mu = 0$, as needed. We then read off the anti-ghost in this gauge
\be 
\bar{c}_\mu=\left(\mathfrak{a}_1+\tfrac{1}{2}\mathfrak{a}_2\right) \bar{c}\circ A_\mu
+\left(\tfrac{1}{2}\mathfrak{a}_2+\mathfrak{a}_3-\tfrac{1}{2}\mathfrak{a}_6 \right)\frac{\partial_\mu}{\partial_0}\bar{c}\circ A_0
-\left(\mathfrak{a}_4-\tfrac{1}{2}\mathfrak{a}_7\right)n_\mu\bar{c}\circ A_0 +\mathfrak{a}_5\frac{\partial_\mu}{\square}\bar{c}\circ\partial A\ .
\ee 
This allows us to proceed to the gauge fixing in gravity
via
\be 
\begin{aligned}
\mathcal{Q} \bar c_\mu&=\frac{1}{\xi}G_\mu[h_{\mu\nu},\varphi]\\
&=\frac{1}{\xi}\Bigg[\left(\mathfrak{a}_1+\tfrac{1}{2}\mathfrak{a}_2\right)A_0\circ A_\mu
+\left(\tfrac{1}{2}\mathfrak{a}_2+\mathfrak{a}_3-\tfrac{1}{2}\mathfrak{a}_6 \right)\frac{\partial_\mu}{\partial_0}A_0\circ A_0
-\left(\mathfrak{a}_4-\tfrac{1}{2}\mathfrak{a}_7\right)n_\mu A_0\circ A_0\\
& \  +\mathfrak{a}_5\frac{\partial_\mu}{\square}A_0\circ \partial A
+\xi\left(\mathfrak{a}_1+\mathfrak{a}_2+\mathfrak{a}_3+\mathfrak{a}_5-\tfrac{1}{2}\mathfrak{a}_6\right)\partial_\mu\left( c\circ \bar{c}\right)
-\xi\left(\mathfrak{a}_4-\tfrac{1}{2}\mathfrak{a}_7 \right)n_\mu\partial_0 \left( c\circ \bar{c}\right)\Bigg]\ .
\end{aligned}
\ee 

Then we require that the BRST gauge-fixing functional for the graviton is also in the temporal form, i.e
\be
G_\mu[h_{\mu\nu},\varphi]=h_{0\mu}\ .
\ee
Using the dictionary \eqref{dict_axial_gauge}, one can show that this is achieved if we impose the following constraints on the parameters:
\begin{eqnarray}
&&\mathfrak{a}_4=-\mathfrak{a}_6\ ,\quad 
\mathfrak{a}_5=0\ ,
\quad \mathfrak{a}_7=2\left(\mathfrak{a}_1+\mathfrak{a}_2+\mathfrak{a}_3\right)-3\mathfrak{a}_6\ , \nonumber\\
&&
\quad\mathfrak{a}_8=0\ ,\quad \mathfrak{a}_9=-\left(\mathfrak{a}_1+\mathfrak{a}_2+\mathfrak{a}_3\right)+\mathfrak{a}_6 \ ,    
\end{eqnarray}
leading to the restricted dictionary
\be \label{dict_axial_gauge_step 1}
\begin{aligned}
h_{\mu\nu}=&\mathfrak{a_1} A_{\mu}\circ A_{\nu} + \mathfrak{a_2}\frac{1}{\partial_0} \partial_{(\mu}A_{\nu)}\circ A_0
+ \mathfrak{a_3}\frac{\partial_\mu\partial_\nu}{(\partial_0)^2}A_0\circ A_0 
+ \mathfrak{a_6}\frac{n_{(\mu}\partial_{\nu)}}{\partial_0} A_0\circ A_0\\
&+\mathfrak{a_6}\xi\frac{\partial_\mu\partial_\nu}{\partial_0}c\circ\bar{c}
- \left[2\left(\mathfrak{a}_1+\mathfrak{a}_2+\mathfrak{a}_3\right)-3\mathfrak{a}_6
\right] \xi n_{(\mu}\partial_{\nu)}c\circ\bar{c}\\
&+\left[\left(\mathfrak{a}_1+\mathfrak{a}_2+\mathfrak{a}_3\right)-\mathfrak{a}_6 \right]\eta_{\mu\nu}\left(\left[A_0\circ A_0+2\xi\, 
\partial_0 c\circ \bar{c}\right]\right)\ ,\\
\varphi=&\mathfrak{b}_1\left[A^\rho\circ A_\rho-\frac{1}{\square}\partial A\circ \partial A\right]-\mathfrak{b}_2\left[A_0\circ A_0+2\xi\, 
\partial_0 c\circ \bar{c}\right]\ .
\end{aligned}
\ee 
We will additionally require temporal gauge in YM to map to temporal  gauge in gravity in the physical sense, i.e.
\be 
A_0=0 \quad \Rightarrow\quad h_{0\mu}=0\ ,
\ee 
which imposes an additional constraint on the parameters,
\be 
\mathfrak{a}_6=2\left(\mathfrak{a}_1+\mathfrak{a}_2+\mathfrak{a}_3\right)\equiv 2\mathfrak{a}_{123} \ ,
\ee 
that further restricts our dictionary to
\be \label{dict_axial_gauge_step 2}
\begin{aligned}
h_{\mu\nu}=&\mathfrak{a_1} A_{\mu}\circ A_{\nu} + \mathfrak{a_2}\frac{1}{\partial_0} \partial_{(\mu}A_{\nu)}\circ A_0
+ \mathfrak{a_3}\frac{\partial_\mu\partial_\nu}{(\partial_0)^2}A_0\circ A_0 
+ 2\mathfrak{a_{123}}\frac{n_{(\mu}\partial_{\nu)}}{\partial_0} A_0\circ A_0\\
&+2\mathfrak{a_{123}}\xi\frac{\partial_\mu\partial_\nu}{\partial_0}c\circ\bar{c}
+4\mathfrak{a_{123}} \xi n_{(\mu}\partial_{\nu)}c\circ\bar{c}
-\mathfrak{a_{123}}\eta_{\mu\nu}\left(\left[A_0\circ A_0+2\xi\, 
\partial_0 c\circ \bar{c}\right]\right)\ ,\\
\varphi=&\mathfrak{b}_1\left[A^\rho\circ A_\rho-\frac{1}{\square}\partial A\circ \partial A\right]-\mathfrak{b}_2\left[A_0\circ A_0+2\xi\, 
\partial_0 c\circ \bar{c}\right]\ .
\end{aligned}
\ee 
Implementing the temporal gauge, we are left with the reduced dictionary  
\be \label{dict_axial_gauge_ij}
\begin{aligned}
h_{ij}=&\mathfrak{a_1} A_{i}\circ A_{j}+2\mathfrak{a_{123}}\xi\frac{\partial_i\partial_j}{\partial_0}c\circ\bar{c}
-2\mathfrak{a_{123}}\delta_{ij}\xi\, 
\partial_0 c\circ \bar{c}\ ,\\
\varphi=&\mathfrak{b}_1\left[A^i\circ A_i-\frac{1}{\partial^i\partial_i}\partial^j A_j\circ \partial^k A_k\right]
-2\mathfrak{b}_2\xi\, 
\partial_0 c\circ \bar{c}\ ,
\end{aligned}
\ee  
where we replaced the $\square$ in the denominator of the dilaton dictionary with the spatial $\partial_i^2$ to preserve the (residual) BRST symmetry after choosing the temporal gauge.

As we will see in the next section, this reduced dictionary prepares us for a straightforward extension to dS backgrounds. Note that if we had chosen the axial gauge, for example, $n^\mu = (0,0,0,1)$, the analysis would have been very similar to the temporal gauge choice. As we will show later, $n^\mu = (0,0,0,1)$ turns out to be helpful for generalization to AdS space.

\section{dS space}\label{sec:ds space}
 After studying the convolutional double copy dictionary in flat space with temporal gauge choice, we are ready to generalize the discussion to de Sitter space. We will work with the conformal metric in the flat slicing coordinate
\begin{eqnarray}
\label{eq,dSmetric}
    d s^2=\left(-\frac{1}{ H \eta }\right)^2\left(-d \eta^2+d \vec{x}^2\right)\ ,
\end{eqnarray}
where $H$ is the Hubble constant denoting the size of the observable universe\footnote{This conformal flat metric can be generalized to the FLRW metric with flat spatial geometry; we leeave this for future study.}. 

We will take the metric perturbation $h_{\mu\nu}$ and the gauge field $A_\mu$ to live on this fixed dS background\footnote{This corresponds to the Type B double copy in the classification of \cite{Bahjat-Abbas:2017htu}.}. The Maxwell action with the temporal gauge fixing term on the de Sitter background is 
\begin{eqnarray}\label{temp_BRST_L}
    \mathcal{S}_A=\int d^4 x \sqrt{-\bar g}\left(-\frac{1}{4} F^{\mu \nu} F_{\mu \nu}+\frac{1}{2 \xi}\left(A_0\right)^2+\bar{c} \nabla_0 c - j_\mu A^\mu - \bar c j_{c} - c j_{\bar c}\right)\ ,
\end{eqnarray}
where $\bar g$ is the background metric for dS space, and $F^{\mu\nu}=\nabla^\mu A^\nu-\nabla^\nu A^\mu$ is the field strength in the curved spacetime. For a conformal metric, the first term in the above action reduces to the flat space expression. However, this is not true for the gauge-fixing and source terms. For simplicity, we use $\nabla$ instead of $\bar\nabla$ to represent the covariant derivative with respect to the background metric.

For the gravitational sector, we use the linearized Einstein-Hilbert action 
\begin{eqnarray}
    \mathcal{S}_{GR} &= &\int d^4 x \sqrt{-\bar g} \bigg(-\frac{1}{4} h^{\mu \nu} E_{\mu \nu}+\frac{1}{2 \xi_{(h)}}\left(h_{0 \mu}\right)^2-\frac{1}{4}(\nabla \varphi)^2+\bar{c}^\mu \nabla_0 c_\mu+\bar{c}_\mu\nabla^\mu c_0   \nonumber\\
    && \qquad \qquad \qquad   - T_{\mu\nu} h^{\mu\nu}  -\bar c^\mu j_{c\mu} - c^\mu j_{\bar c \mu}\bigg)\ ,
\end{eqnarray}
where $\varphi$ is the dilaton, and $E_{\mu\nu}$ is the linearized Einstein tensor that takes the form
\begin{eqnarray}
E_{\mu\nu} &=& \frac{1}{2}\left(-\bar{R} h_{\mu\nu} - \nabla_\mu\nabla_\nu h + \bar{g}^{\alpha\beta}\left( \nabla_\alpha\nabla_\mu h_{\nu\beta }+\nabla_\alpha\nabla_\nu h_{\mu\beta }  \right) \right.\nonumber\\
&& \left. - \nabla^2 h_{\mu\nu} + \bar{g}_{\mu\nu}\left(h^{\alpha\beta}\bar{R}_{\alpha\beta} -\nabla_\alpha\nabla_\beta h^{\alpha\beta } +\nabla^2 h  \right)\right)   \ .    
\end{eqnarray}
$\bar{R} = 12 H^2$ is the Ricci curvature in 4D dS background, $\frac{1}{2 \xi_{(h)}}(h_{0\mu})^2$ is the temporal gauge fixing term, and $c_\mu, \bar c_\mu$ are the corresponding graviton ghost and anti-ghost, respectively.   

We now wish to construct a convolutional double copy appropriate for dS space. We first notice that, as denoted by the scale factor in the metric, the time translational symmetry is broken. To describe the gauge fields and metric perturbations in dS space, we can only apply the Fourier transformation on the spatial directions. The temporal coordinate will have to be treated differently. To best capture the scale symmetry in dS space, we apply the Mellin transformation, 
\begin{eqnarray}
     h(\eta)=\int \frac{d s}{2 \pi i} 2 \tilde{h}(s) \eta^{-2 s+\frac{d}{2}}\ , \quad \tilde{h}(s)=\int_0^{\infty} d \eta ~ h(\eta) \eta^{2 s-\frac{d}{2}-1}\ . 
\end{eqnarray}
One can see that $\eta^{-2s+ \frac{d}{2}}$ captures the scale symmetry $h(\lambda\eta)= \lambda^{\Delta}h(\eta)$, where $\Delta$ is the scaling dimension. A nice comparison between the Mellin transform and the Fourier transform can be found in \cite{Sleight:2021plv}. 
 
To obtain the dictionary in dS space, we combine the standard convolution with a convolution in Mellin space \eqref{eq,convoMellin} as follows
\begin{eqnarray}\label{mixd_conv}
   h(\eta,x)=  [f \star g](\eta, x)=\int \frac{d \tau}{\tau} \int d^3 y f(1\times \tau, 0+x) g\left(\frac{\eta}{\tau}, y-x\right) \ .
\end{eqnarray}
Here the unit $1$ in the time direction indicates the scaling symmetry, and the unit $0$ in the spatial directions indicates the translational symmetry. 
One can easily check that in the Mellin-Fourier space, the convolution defined above is simply the product,
\begin{eqnarray}
  \tilde h(s,k)=   \tilde f(s,k)\tilde g(s,k)\ .
\end{eqnarray}
The convolutional double copy can be defined similarly through the $\circ$ product,
\begin{eqnarray}
\label{eq,convolutionindS}
    A_\mu \circ A_\nu \equiv A_\mu^a \star\left(\Phi^{-1}\right)^{a a^{\prime}} \star A_\nu^{a^{\prime}} \ ,
\end{eqnarray}
with the spectator field $\Phi$ serving as the kernel.

An important requirement on our dictionary is that it reproduces the correct gravity BRST transformations 
\begin{equation}
\mathcal{Q}h_{\mu\nu}=\nabla_{\mu}c_{\nu}+\nabla_\nu c_\mu,\quad
\mathcal{Q}\varphi=0\ ,
\label{gravBRSTcurved}
\end{equation}
from the YM transformations:
\begin{equation}
     \mathcal{Q} A_\mu = \nabla_\mu c,\ \mathcal{Q} c = 0, \ \mathcal{Q} \bar c
     = \frac{1}{\xi} n A 
     = \frac{1}{\xi} A_0 \ .
 \end{equation} 

 Notice that the covariant derivative does not have the nice flat space property, Eq.\eqref{derivrule}, since only the spatial partial derivatives pass through the convolution. This poses an apparent issue for a standard convolutional dictionary satisfying \eqref{derivrule}, since a term of the form $A_\mu\circ A_\nu$ will transform as
\be 
\mathcal{Q}(A_\mu\circ A_\nu)=\partial_\mu c\circ A_\nu+A_\mu\circ\partial_\nu c=\partial_\mu \left(c\circ A_\nu\right) + \partial_\nu \left(A_\mu\circ c\right) \ ,
\ee 
and it is unclear how to reproduce the full covariant derivative in \eqref{gravBRSTcurved}. To find the dictionary in de Sitter space, we need to resolve the mixing between the temporal components and the spatial ones as inherited from the covariant derivative,
\begin{eqnarray}
    \nabla_0 A_0 = \partial_0 A_0-\frac{1}{\eta}A_0 \ , \nabla_i A_0 = \partial_i A_0 + \frac{1}{\eta} A_i\ ,\nabla_0 A_i = \partial_0 A_i + \frac{1}{\eta} A_i\ , \nabla_i A_j = \partial_i A_j + \frac{1}{\eta} A_0\ . 
\end{eqnarray}

It is straightforward to see that restricting to fields satisfying temporal gauge $A_0 = 0$ eliminates the mixing between temporal components with the spatial components when taking the covariant derivative, and the YM BRST residual transformations essentially reduce to flat ones
\begin{equation}\label{axial gauge YM transf}
     \mathcal{Q} A_i = \partial_i c,\ \mathcal{Q} c = 0, \ \mathcal{Q} \bar c
     = 0 \ .
 \end{equation}
Then, choosing the temporal gauge for gravity, $h_{0\mu}=0 $, together with $c_0=0$, we see that \eqref{gravBRSTcurved} reduces to 
\be 
\mathcal{Q}h_{ij}=\partial_{i}c_{j}+\partial_j c_i,\quad
\mathcal{Q}\varphi=0,
\ee 
which follows from \eqref{axial gauge YM transf}.  

The flat space dictionary in temporal gauge, Eq.\eqref{dict_axial_gauge_ij}, can be directly extended to dS space with the modified convolution law given in Eq.\eqref{eq,convolutionindS} and \eqref{mixd_conv}\footnote{ Let us comment on the term $\partial_0 c$ appearing in the dilaton dictionary. In order to preserve the gauge choice $A_0=0$, we expect this to vanish everywhere except perhaps in a localised region of spacetime. This can be achieved by an appropriate choice of source $j_c$ in \eqref{temp_BRST_L}. This is analogous to the situation in Lorenz gauge, where we have $\square c\propto \delta^3(\vec x)$. However, such terms are not negligible from the perspective of the convolution, which integrates over the whole of spacetime, which explains its appearance in the dictionary.}, 
\begin{eqnarray}
\label{eq,dSdictionary}
    \begin{aligned}
h_{ij}=&\mathfrak{a_1} A_{i}\circ A_{j}+2\mathfrak{a_{123}}\xi\frac{\partial_i\partial_j}{\partial_0}c\circ\bar{c}
-2\mathfrak{a_{123}}\delta_{ij}\xi\, 
\partial_0 c\circ \bar{c}\ ,\\ 
\varphi=&\mathfrak{b}_1 H^2\eta^2 \left(\delta_{ij}\left[ A_i \circ A_j \right]-   \frac{1}{\partial_i\partial_i }   \left[ \partial_k A_k \circ \partial_m A_m \right]  \right) 
-2\mathfrak{b}_2\xi\, 
\partial_0 c\circ \bar{c}\ ,
\end{aligned}
\end{eqnarray}
where the repeated indices indicate the summation over spatial dimensions. We will later apply this dictionary to physical solutions in \autoref{sec:solution}, and show that it correctly generalises the known flat-space results.

To conclude this section, we briefly comment on the generalisation to AdS space. The AdS spacetime is also conformally flat,
 \begin{eqnarray}
     ds^2_{\text{AdS}} = \frac{1}{z^2} (-d\eta^2 + d\vec x^2 )\ ,
 \end{eqnarray}
 with the conformal factor being a function of radial direction $z$. Here we will consider a different constant vector in the gauge fixing term, $n^\mu = (0,0,0,1)$, which sets $A_3=0$. Following the same process, we obtain the following dictionary 
 \begin{eqnarray}
\label{eq,AdSdictionary}
    \begin{aligned}
h_{ij}=&\mathfrak{a_1} A_{i}\circ A_{j}+2\mathfrak{a_{123}}\xi\frac{\partial_i\partial_j}{\partial_z}c\circ\bar{c}
-2\mathfrak{a_{123}}\delta_{ij}\xi\, 
\partial_z c\circ \bar{c}\ ,\\ 
\varphi=&\mathfrak{b}_1 H^2 z^2 \left(\eta_{ij}\left[ A_i \circ A_j \right]-   \frac{1}{\eta\partial_i\partial_i }  \eta_{kl}\eta_{mn}\left[ \partial_k A_l \circ \partial_m A_n \right]  \right) 
-2\mathfrak{b}_2\xi\, 
\partial_z c\circ \bar{c}\ ,
\end{aligned}
\end{eqnarray} 
where now $i,j = 0,1,2$, and $\eta_{ij} = -1$ when $i = j = 0$, $\eta_{ij} = 1$ when $i = j = 1,2$, and vanishes otherwise. 

We will now study some examples of physical solutions, in particular the (dS-)Schwarzschild metric and the point-like charge in gauge theory.

\section{solution}
\label{sec:solution}
\subsection{flat space Schwarzschild}
In this subsection, we discuss a specific example of the convolution double copy in flat space. We will relate the point-source Maxwell solution with the Schwarzschild BH solution. This has already been done in \cite{Luna:2020adi} in Lorenz and de Donder gauge, respectively. Here, we will re-do the calculation in temporal gauge, in order to set the stage for the Schwarzschild solution in a de Sitter background, which will be explored in the next subsection. 

We start with the linearised YM field for the point charge solution,
\begin{eqnarray}
\label{eq,gaugecharge}
    A_\mu^a=\frac{g \alpha^a}{4 \pi r}\left(1,0,0,0\right) \ ,
\end{eqnarray}
with $\alpha^a$ a normalised constant color vector. We can put this in temporal gauge by applying the gauge transformation in Eq.\eqref{eq,gaugespatial}, to get 
\begin{eqnarray}\label{YMtemporal}
   \left(A_0^{a}\right)^\prime = 0\ , \quad  \left(A_i^{a}\right)^\prime=\frac{g t}{4 \pi} \frac{x_i}{r^3} \alpha^a \ .
\end{eqnarray}
We will also need the spectator field
\be \label{spec_pos}
\Phi^{aa'}=\frac{g\delta^{aa'}}{r}\ ,
\ee 
for the kernel of the convolutional double copy.

On the other hand, the Schwarzchild BH solution takes the form 
\begin{eqnarray}
    ds^2 = -\left(1-\frac{r_s}{r}\right)dt^2 +\left(1-\frac{r_s}{r} \right)^{-1} dr^2 + r^2 d\theta^2 + r^2 \sin^2\theta d\varphi^2 \ ,
\end{eqnarray}
where $r_s$ is the Schwarzchild radius of the black hole. 
We do the coordinate transform to the isotropic coordinates and find the metric perturbation takes the form,
\begin{eqnarray}
\label{eq,gravitonschwarz}
    h_{00} = \frac{r_s}{r}\ , h_{0i}=0\ , h_{ij} = \frac{r_s}{r} \delta_{ij}\ ,
\end{eqnarray}
where $r =\sqrt{x^2 + y^2 + z^2}$ is the radius outside the horizon \cite{dyson_1925}. 
 
We now go to temporal gauge, $h_{0\mu}=0$, see details in \autoref{app,axialgauge}.  The spatial graviton in this gauge takes the form
\begin{eqnarray}
\label{eq,gravitonaxial}
    h_{i j}^{\prime}=\frac{r_s}{r }\delta_{i j} -\frac{t^2 r_s }{2  } \left(\frac{ \delta_{i j}}{r^3}-\frac{3   x_i x_j}{r^5}\right) -  t^2 r_s\frac{\delta_{ij}}{3} 2\pi \delta^3 (\vec x)  
    \ ,
\end{eqnarray} 
 We will neglect the notation  $\prime$ in the gauge field and graviton field from now on.

We would like to find the dictionary for this physical solution, both for the graviton field and for the vanishing dilaton field that does not show explicitly in the Schwarzschild solution. To compute the dilaton dictionary, we first notice the Fourier transform of the gauge field takes the form,
\begin{eqnarray}
    \mathcal{F}(A_i^a ) = g \pi \frac{\delta(k_0)}{k_0} \frac{k_i}{k^2}\alpha^a\ ,
\end{eqnarray}
with $\alpha^a$ a normalised constant color vector as before. Therefore, the convolution $A_i \circ A_j$ takes the form
\begin{eqnarray}
\label{eq,AicircAj}
    A_i \circ A_j = \frac{g \pi}{4} \frac{\delta(k_0)}{k_0^2}   \frac{k_i k_j}{k^2} \ ,
\end{eqnarray}
where we used the fact that the Fourier transform of the spectator scalar field is $g \delta(k_0) \frac{4\pi}{k^2}\delta^{aa'}$. Then the first two terms in the dilaton dictionary can be obtained from the above expressions, 
\begin{eqnarray}
    A^i \circ A_i - \frac{1}{\partial^i\partial_i} \partial^j A_j \circ \partial^k A_k = \frac{g \pi }{4}\frac{\delta(k_0)}{k_0^2 }   + \frac{1}{  - k^2} \frac{g \pi }{4}\frac{\delta(k_0)}{k_0^2 }  k^2 = 0   \ .
\end{eqnarray}
Then we set the coefficient of the ghost-antighost term in the dilaton dictionary in \eqref{eq,dSdictionary} to vanish in order to get a zero dilaton, i.e. $\mathfrak{b}_2 = 0$. 

To map with the Fourier transformation of the graviton 
\begin{eqnarray}
\label{eq,gravitonfourier}
    \mathcal{F}(h_{ij}) = r_s 8\pi^2 \frac{\delta(k_0)}{k_0^2} \frac{k_i k_j}{k^2} + r_s 8\pi^2 \frac{\delta_{ij }\delta(k_0)}{k^2} \ ,
\end{eqnarray}
we need a ghost-antighost term which behaves like 
\begin{eqnarray}
    \xi \partial_0 c \circ \bar c =  \frac{g \pi }{4}\frac{\delta(k_0)}{ -k^2  } \ .  
\end{eqnarray}
Plugging it back in the graviton dictionary, we have
\begin{eqnarray}
\mathfrak{a}_1  A_i \circ A_j   + 2 \mathfrak{a}_{123} \left(\frac{\partial_i \partial j}{\partial_0^2 } - \delta_{ij } \right) \xi \partial_0 c \circ \bar c = \left(\mathfrak{a}_1  -  \mathfrak{a}_{123}   \right) \frac{g \pi}{4} \frac{\delta(k_0)}{k_0^2} \frac{k_i k_j}{k^2} +  \mathfrak{a}_{123}    \frac{g \pi}{4} \frac{\delta_{ij}\delta(k_0)}{k^2} \ .
\end{eqnarray}

 We therefore require the following mapping between the coefficients
\begin{eqnarray}
   \frac{g \pi}{4}  \mathfrak{a}_{123}  = 8\pi^2 r_s\ , \quad   \frac{g \pi}{4} \mathfrak{a}_{1} = 16 \pi^2 r_s\ ,
\end{eqnarray}
in order to reproduce the Schwarzchild solution. The details of the various Fourier transformations used have been saved for Appendix \eqref{app,FT}. For flat space, one can also start from the full dictionary,  Eq.\eqref{dict_axial_gauge_step 2},  in the absence of physical gauge fixing.  Imposing the vanishing of the dilaton will in this case result in a non-trivial ghost contribution, as expected. We present the corresponding calculation in Appendix \eqref{app,dilaton}.
 
 We conclude with a remark on the spectator scalar \eqref{spec_pos}. We note that the convolution inverse of this spectator should not be thought of as a stand-alone object (indeed it would be divergent), but rather as a formal operator acting on one of the $A_i$ factors.   
 Note that the $\delta(k_0)$ in the Fourier transform of the spectator scalar has been chosen exactly as to cancel the $\delta(k_0)$ dependence in one of the $A_i$ factors in the product $A_i \circ A_j \equiv A_i^a \star\left(\Phi^{-1}\right)^{a a^{\prime}}\star A_j^{a^{\prime}}$, yielding a finite final result by construction. Alternatively, in certain cases one can circumvent the issue and have a mathematically well defined spectator inverse by choosing both the spectator and one of the YM factors to be of the form of a full 4-dimensional delta function $\delta^4(x)$ (which is its own convolution inverse)\cite{Godazgar:2022gfw,Cardoso:2016amd,Cardoso:2016ngt}\footnote{For further remarks on circumventing analytical issues that might arise in the convolution product see also \cite{Borsten:2023ned}.}. Here we have chosen both YM factors to be of the form \eqref{eq,gaugecharge}-\eqref{YMtemporal}, in order to connect with other double copy proposals for classical solutions in the literature, which inevitably leads to a spectator of the form \eqref{spec_pos}, and the formal operator interpretation.

\subsection{dS-Schwarzschild}
We now want to seek a solution for de Sitter space. The linearised YM equation in a dS background with a point source is
\begin{eqnarray}
    \nabla _\mu F^{\mu \nu a} = g  \delta^{(3)}(\vec{x}) \delta_0^\nu \alpha^a \ ,
\end{eqnarray}
and the solution in temporal gauge is 
\begin{eqnarray}
    A_0^a = 0,\ A_i^a = \frac{g x_i (-\eta)}{r^3 }\alpha^a =  - g (-\eta) \partial_i \left(\frac{1}{r}\right)\alpha^a \ .
\end{eqnarray}

For the gravitational sector, we start from the dS-Schwartzchild solution in the static chart\footnote{For the classical Kerr-Schild double copy for this solution see \cite{Carrillo-Gonzalez:2017iyj,Bahjat-Abbas:2017htu}.}, 
\begin{eqnarray}
    ds^2 = -\left(1-\frac{r_s}{r}-\frac{\Lambda r^2}{3}\right) \mathrm{d} t^2+\left(1-\frac{r_s }{r}-\frac{\Lambda r^2}{3}\right)^{-1} \mathrm{~d} r^2+r^2 \mathrm{~d} \Omega^2 \ ,
\end{eqnarray}
where $\Lambda$ is the cosmological constant that relates to the Hubble constant via $\Lambda/3 = H^2$, and $r_s$ is the Schwarzschild radius of the black hole. We then expand the metric to $\mathcal{O}(r_s)$ and apply the coordinate transformation to the flat slicing coordinate. The metric turns to
\begin{eqnarray}
    ds^2& =& \frac{1}{H^2\eta^2}\left(-d \eta^2 + d\vec x^2\right) - \frac{r_s}{r} \frac{d\eta^2}{H \eta} \frac{\left(1+ \frac{r^2}{\eta^2 }\right)}{\left(1- \frac{r^2}{\eta^2 }\right)^2} + \frac{4r_s }{r} \frac{x_i dx^i d\eta}{H \eta^2} \left(1- \frac{r^2}{\eta^2 }\right)^{-2} \nonumber\\
    && - \frac{r_s }{r^3 H \eta } \frac{\left(1+ \frac{r^2}{\eta^2 }\right)}{\left( 1- \frac{r^2}{\eta^2 }\right)^2} (x_i dx^i)^2 \nonumber \ ,
\end{eqnarray}
and one can read off the graviton from the above expression.  
Following \autoref{app,axialgauge}, we then go to temporal gauge $h_{0\mu}=0$, where the new spatial $h_{ij}^\prime$ takes the form 
% \red{remove first line ?},
\begin{eqnarray}
\label{eq,gravitondS}
 h_{ij}^\prime 
 &=& \frac{r_s x_i x_j (-\eta)}{H r^5 } - \frac{r_s \delta_{ij} (-\eta)}{3H  r^3 }   +4\pi \frac{r_s \eta}{9H}\delta_{ij}\delta^3 (\vec x) \\
 &=& \frac{r_s (-\eta)}{3 H} \partial_i\partial_j \left(\frac{1}{r}\right)\ .
\end{eqnarray}
Notice the $\delta^3(\vec x)$ function is coming from the Laplacian operator
\begin{eqnarray}
   \sum_i \partial_i\partial_i\left(\frac{1}{r} \right) = - 4\pi \delta^3(\vec x)\ ,
\end{eqnarray}
which is essential to capture the trace of graviton.

The gauge field and graviton field in Mellin space and Fourier space take the form
\begin{eqnarray}
    \tilde A_i^a (s,k)  &=&  \int_0^\infty d(-\eta) \int d^3 \vec x (-  g )\partial_i \frac{(-\eta ) }{r } e^{i\vec k\cdot\vec x} (-\eta)^{2s -\frac{d}{2} -1 }\alpha^a \nonumber\\
    &=& 4\pi g \frac{k_i}{k^2} 2\pi \delta\left(2s - \frac{d}{2} +1 \right)\alpha^a\ ,
\end{eqnarray}
\begin{eqnarray}
\label{eq,Mellingrav}
\tilde h_{ij}(s,k ) = - \frac{r_s}{3H} 2\pi i \delta\left(2s - \frac{d}{2} +1 \right) \frac{4\pi k_i k_j }{k^2}\ .
\end{eqnarray}
 
With a spectator field
\begin{eqnarray}
    \Phi^{aa'}(\eta ,x ) = -\eta \frac{\delta^{aa'}}{r} \ , \tilde \Phi^{aa'} (s,k) = \frac{4\pi}{k^2} 2\pi i \left(2s - \frac{d}{2} +1 \right)\delta^{aa'} \ ,
\end{eqnarray}
the convolution becomes 
\begin{eqnarray}
  A_i\circ A_j =  -  2\pi i \delta \left(2s - \frac{d}{2}+1 \right)   g 4\pi \frac{k_i k_j}{k^2} \ ,
\end{eqnarray}
which matches our graviton in Mellin/Fourier space obtained in Eq.\eqref{eq,Mellingrav}. Comparing to the general dictionary \eqref{eq,dSdictionary}, we see that the double copy works with a vanishing ghost term. The convolution therefore takes a particularly simple form, 
\begin{eqnarray}
    h_{i j}=\mathfrak{a}_1 A_i \circ A_j \ .
\end{eqnarray}
And one can check that the dilaton field automatically vanishes with $\mathfrak{b}_2 = 0$. 

It is remarkable that we do not need the ghost to contribute in the convolutional dictionary in de Sitter space, which indicates that the temporal gauge is the proper gauge choice for de Sitter.

\section{Discussion and Conclusion}
In this paper, we have shown that the convolutional double copy can be extended to (anti) de Sitter backgrounds. We found that arbitrary gravitational fluctuations in temporal gauge, as well as their residual symmeteries, can be constructed from counterparts in the YM theory. As an example, we demonstrated how the Schwarzschild solution in a dS background follows from an electromagnetic potential sourced by a point charge. Encouragingly, this agrees with the known results in the flat space limit. 

We have found the temporal gauge to be necessary in formulating the simplest extension of the flat-space convolution to de Sitter. Of course, it would be interesting to construct a more general convolution product on an (A)dS background, which could deal with arbitrary gauge choices.  

On a more practical note, it would also be worthwhile to explore more general spacetimes of interest to cosmology. For example, we note that the subset of FLRW metrics with vanishing scalar curvature of the 3-space is conformally flat. The conformal factor is a simple generalisation of that in Eq.\eqref{eq,dSmetric}, again depending only on the time coordinate, thus making it a good candidate for a similar double copy construction in the temporal gauge. It is known that the  non-spinning black hole solution in a general conformal spacetime can be described by the McVittie metric\cite{1933MNRAS..93..325M}, and it would be very interesting to reproduce this via the double copy.

Another direction for generalising our work is to go beyond the point like charge sources to higher multipole moments and study the corresponding gauge/gravity response. Additionaly, as (A)dS spacetime has an interesting asymptotic behavior, it would be of interest to generalize the results obtained in \cite{Campiglia:2021srh,Adamo:2021dfg} for flat spacetime. Finally, in flat space, there has been progress in understanding how double copy formulations for fields and classical solutions connect with the amplitudes formulation.
Another possible direction would then be to connect with the correlator double copy programme \cite{Mei:2023jkb,Armstrong:2020woi,Albayrak:2020fyp,Alday:2021odx,Diwakar:2021juk,Sivaramakrishnan:2021srm,Cheung:2022pdk,Herderschee:2022ntr,Drummond:2022dxd,Farrow:2018yni,Lipstein:2019mpu,Jain:2021qcl,Zhou:2021gnu,Armstrong:2022csc,Lipstein:2023pih}, which we leave for future work.

\acknowledgements
We thank Mariana Carillo-Gonzales and Sam S.C. Wong for useful discussions. We thank KITP for hosting us during the initial discussions that led to the paper. The work of QL is supported in part by US Department of Energy (HEP) Award DE-SC0013528, and in part by World Premier International Research Center Initiative (WPI), MEXT, Japan. The work of SN is supported in part by STFC consolidated grant T000708.

\appendix
\section{Fourier Transformation}
\label{app,FT}
In this section, we list some useful integral formulas following the conventions in \cite{Smirnov:2012gma}. The expressions involving vector indices are derived by taking the derivative with respect to the spatial coordinate,
\begin{eqnarray}
    \begin{aligned} \int \frac{d^d \vec{k}}{(2 \pi)^d} \frac{e^{i \vec{k} \cdot \vec{x}}}{\left(\vec{k}^2\right)^\alpha} & =\frac{1}{(4 \pi)^{d / 2}} \frac{\Gamma(d / 2-\alpha)}{\Gamma(\alpha)}\left(\frac{\vec{r}^2}{4}\right)^{\alpha-d / 2} \\ \int \frac{d^d \vec{k}}{(2 \pi)^d} \frac{k_i}{\left(\vec{k}^2\right)^\alpha} e^{i \vec{k} \cdot \vec{x}} & =i x_i \frac{\Gamma(d / 2-\alpha+1)}{2(4 \pi)^{d / 2} \Gamma(\alpha)}\left(\frac{\vec{r}^2}{4}\right)^{\alpha-d / 2-1} \\ \int \frac{d^d \vec{k}}{(2 \pi)^d} \frac{k_i k_j}{\left(\vec{k}^2\right)^\alpha} e^{i \vec{k} \cdot \vec{r}} & =\frac{\Gamma(d / 2-\alpha+1)}{(4 \pi)^{d / 2} \Gamma(\alpha)}\left(\frac{\delta_{i j}}{2}+(\alpha-d / 2-1) \frac{x_i x_j}{\vec{r}^2}\right)\left(\frac{\vec{r}^2}{4}\right)^{\alpha-d / 2-1}\end{aligned}\ ,
\end{eqnarray}
where the second equation is obtained by taking derivative $-i \partial_i = -i  \frac{\partial}{\partial x^i }$ on both sides of the first equation, and similarly for the third equation. It is worth pointing out that the above standard Fourier transforms only hold when we are away from the origin, $r= 0$. As is well known, the Laplacian acting on the $1/r$ gives us the $\delta $ function, which is zero everywhere away from the origin. When acting with derivatives $- \partial_i \partial_j $ on the first equation, one should also recover the $\delta$ -behavior when taking the trace. This leads to a special formula that we will use in Eq.\eqref{eq,gravitonfourier},
\begin{eqnarray}
\label{eq,Fourierkikj}
    \int \frac{d^3 \vec k}{(2\pi)^3} \frac{k_i k_j}{k^2} e^{i \vec k \cdot \vec r} = \frac{1}{4\pi} \left( \frac{\delta_{ij}}{r^3} - \frac{3x_i x_j }{r^5}  \right) + \frac{\delta_{ij}}{3} \delta^3 (\vec x) \ ,
\end{eqnarray}
where the $\delta^3 (\vec x )$ function can also be moved to the left hand side as a constant. 

The inverse Fourier transformation takes the form,
\begin{eqnarray}
    \begin{aligned} \int  d^d \vec{x }  \frac{e^{- i \vec k  \cdot \vec x}}{\left(\vec{x}^2\right)^\alpha} & =\pi^{d / 2} \frac{\Gamma(d / 2-\alpha)}{\Gamma(\alpha)}\left(\frac{\vec{k}^2}{4}\right)^{\alpha-d / 2} \\ \int d^d \vec{x}\frac{\vec{x}_i}{\left(\vec{x}^2\right)^\alpha} e^{-i \vec x  \cdot \vec k} & =i k_i \pi^{d / 2} \frac{\Gamma(d / 2-\alpha+1)}{2 \Gamma(\alpha)}\left(\frac{\vec{k}^2}{4}\right)^{\alpha-d / 2-1} \\ \int  d^d \vec x \frac{\vec x_i \vec x_j}{\left(\vec x^2\right)^\alpha} e^{- i \vec x \cdot \vec k} & =\pi^{d / 2} \frac{\Gamma(d / 2-\alpha+1)}{ \Gamma(\alpha)}\left(\frac{\delta_{i j}}{2}+(\alpha-d / 2-1) \frac{k_i k_j}{\vec k^2}\right)\left(\frac{\vec k^2}{4}\right)^{\alpha-d / 2-1}\end{aligned}\ .
\end{eqnarray}

As to the Fourier transformations involving time, we treat them in the following way,
\begin{eqnarray}
    f(t)= t^2 \ , f(t) = \int \frac{d k_0}{2\pi} e^{ i k_0 t} \tilde f(k_0)\ ,
\end{eqnarray}
inversely,
\begin{eqnarray}
  \tilde f(k_0) =  \int d t e^{ - i k_0 t} f(t)\ .
\end{eqnarray}

We have
\begin{eqnarray}
 2 =  \frac{d^2}{d t^2 }  f(t) = \int \frac{d k_0}{2\pi}  \frac{d^2}{d t^2 } e^{ i k_0 t}  \tilde f(k_0) = - \int \frac{d k_0}{2\pi}  k_0 ^2 e^{ i k_0 t}   \tilde f(k_0)  \ .
\end{eqnarray}
Since the constant under the Fourier transformation gives us the $\delta$ function, we can express $\tilde f(k_0)$ as $- 4\pi \delta(k_0)/k_0^2$. Though this expression is not well-defined as a distribution function, one should think of it from the perspective of the derivative operation. Similarly. for $f(t) = t$, we have
\begin{eqnarray}
    1 = \frac{d}{d t }  f(t) = \int \frac{dk_0}{2\pi} \frac{d}{d t } e^{ i k_0 t}  \tilde f(k_0) = i \int \frac{dk_0}{2\pi} k_0  e^{ i k_0 t}   \tilde f(k_0) \ ,
\end{eqnarray}
Its Fourier transformation is $2\pi \delta(k_0)/(i k_0) $. 

\section{Temporal Gauge }
\label{app,axialgauge}
\subsection{flat space}
In this section, we show how to apply gauge transformations in order to write our solutions in temporal gauge. We start with the flat space. Here we will leave the gauge index implicit, as it does not affect the calculations at linear level.

We start from the solution in Lorenz gauge Eq.\eqref{eq,gaugecharge}, and perform the following gauge transformation,
\begin{eqnarray}
\label{eq,gaugespatial}
    A_0^\prime = A_0+\partial_0 \Lambda = 0,\ \Lambda = - \frac{g~ t} {4\pi r}+ f(r), \ A_i^\prime = \frac{g~ t  }{4\pi  } \frac{ x_i }{  r^3}  +f^\prime(r) \frac{x_i}{r}\ .
\end{eqnarray}
where $f(r)$ denotes the redundant gauge degree of freedom that will be set to zero for simplicity here, and $'$ denotes the derivative with respect to $r$. 

In flat space, the gravitational perturbation transforms under the linearized diffeomorphism through
\begin{eqnarray}
    h_{\mu\nu}^\prime =  h_{\mu\nu} + \partial_\mu\xi_\nu+\partial_\nu\xi_\mu  \ .
\end{eqnarray}
To go to the temporal gauge where $h_{0\mu}^\prime $ vanishes, we first apply the transformation on the temporal component such that
\begin{eqnarray}
    h_{00}^\prime = h_{00}+ 2\partial_0 \xi_0 = 0\ .
\end{eqnarray}
This is gives a differential equation for the gauge parameter $\xi_0$, with solution 
\begin{eqnarray}
    \xi_0 = -\frac{1}{2} \int d t~  h_{00} + F(r) = -\frac{1}{2} \frac{r_s}{r} t + F(r) \ ,
\end{eqnarray}
where $F(r)$ as a function of spatial coordinates serves as the integration constant, and represents the redundant gauge. The next step is to transform $h^\prime_{0i}$ to zero, 
\begin{eqnarray}
    h_{0i}^\prime = h_{0 i} + \partial_0 \xi_i + \partial_i \xi_0 = 0 \ ,   
\end{eqnarray}
which gives 
\begin{eqnarray}
    \xi_i = - \int dt~ h_{0i} - \int dt~ \partial_i \xi_0 = -\frac{t^2}{4} \frac{r_s x_i}{r^3} - t F^\prime (r) \frac{x_i }{r}+ F_i (r)\ ,
\end{eqnarray}
where the $F_i(r)$ are functions of $r$ and serve as the redundant gauge in spatial components. This gives the spatial graviton after the gauge transform, 
\begin{eqnarray}
    h_{i j}^{\prime}=\frac{r_s}{r }\delta_{i j} -\frac{t^2 r_s }{2  } \left(\frac{ \delta_{i j}}{r^3}-\frac{3   x_i x_j}{r^5}\right) \ ,
\end{eqnarray} 
where we have set the redundant gauges $F(r)$ and $F_i(r)$ to zero. 

\subsection{de Sitter space}
In de Sitter space, the diffeomorphism involves the covariant derivatives,
\begin{eqnarray}
    h_{ \mu\nu}\to h^\prime_{\mu\nu} + \nabla_\mu \xi_\nu +\nabla_\nu \xi_\mu   \ .
\end{eqnarray}
\begin{eqnarray}
\delta h_{00}=2 \partial_0 \xi_0+\frac{2}{\eta} \xi_0, \delta h_{0 i}=\partial_0 \xi_i+\partial_i \xi_0+\frac{2}{\eta} \xi_i, \delta h_{i j}=\partial_i \xi_j+\partial_j \xi_i+ \frac{2 \delta_{i j}}{\eta} \xi_0\ .
\end{eqnarray}
Now we want $h^\prime_{00} =0 $, which means 
\begin{eqnarray}
   2\partial_0 \xi_0 + \frac{2}{\eta} \xi_0=  \delta h_{00} = -h_{00} = \frac{r_s }{r} \frac{1}{H \eta} \frac{\left(1+ \frac{r^2}{\eta^2 }\right)}{\left( 1- \frac{r^2}{\eta^2 }\right)^2}  \ , \nonumber\\
   \xi_0 = \frac{r_s}{2 H r}\left(1+ \frac{r^2}{r^2 - \eta^2}  \right) -  \frac{r_s}{H \eta} \tanh^{-1}\frac{\eta}{r}+ \frac{F(x)}{\eta}\ .
\end{eqnarray}

The next step is to make $h_{0i}^\prime = 0 $, this implies that
\begin{eqnarray}
    \frac{2 r_s }{r} \frac{x_i}{H \eta^2}\left(1- \frac{r^2}{\eta^2}\right)^{-2} + \partial_0 \xi_i + \frac{2}{\eta}\xi_i + \partial_i \xi_0 = 0   \ ,
\end{eqnarray}
\begin{eqnarray}
    \xi_i = \frac{ r_s x_i}{2 r (-H\eta)  } - \frac{ r_s x_i (-\eta)  }{6H r^3  }- \frac{ r_s x_i r}{ 2(-H \eta) (\eta^2-r^2) }    +\frac{ r_s x_i}{  H \eta^2} \tanh^{-1}\frac{\eta}{r}  -\frac{x_i F'(r)}{2r } + \frac{F_i(r)}{\eta^2} \ .
\end{eqnarray}
With the $\xi_\mu$ obtained above, we get 
\begin{eqnarray}
    h_{ij}' = h_{ij}+ \partial_i \xi_j + \partial_j \xi_i =\frac{r_s x_i x_j(-\eta)}{H r^5}-\frac{r_s \delta_{i j}(-\eta)}{3 H r^3}+4 \pi \frac{r_s \eta}{9 H} \delta_{i j} \delta^3(\vec{x})\ , 
\end{eqnarray}
which are the new spatial components in temporal gauge used in Eq.\eqref{eq,gravitondS}. 

One can check that the above solutions in temporal gauge satisfy the equations of motion of gauge fields and gravitational fields, 
\begin{eqnarray}
\label{eq,eomAmu}
   && H^2 \eta^2\left(\partial_0^2 A_i-\partial_j^2 A_i+\sum_{j=1}^3 \partial_j \partial_i A_j\right)=j_i\ , \nonumber\\
    && H^2 \eta^2 \partial_0 \partial_i A_i=j_0 \ ,
\end{eqnarray}
where the second line should be treated as a constraint equation from the EoM of the temporal component $A_0$. 

In a similar way, we can obtain the EoM for the graviton as 
\begin{eqnarray}
    \frac{1}{2}\left(\partial^2 h-\partial_i \partial_j h_{i j}\right)+\frac{2}{\eta^2} h+\frac{1}{\eta} \partial_0 h=\frac{T_{00}}{H^2 \eta^2}\ ,
\end{eqnarray}
where $h \equiv \sum_{k= 1,2,3}h_{kk}$ is the trace of the graviton. For $h_{0i}$, the EoM is
\begin{eqnarray}
    \frac{1}{2}\partial_0\left(\partial_i h - \partial_j h_{ij} \right) + \frac{1}{\eta} \left(\partial_i h - \partial_j h_{ij}  \right) = \frac{T_{0i}}{H^2 \eta^2 }\ .
\end{eqnarray}
For the spatial part $h_{ij}$, the EoM is 
\begin{eqnarray}
 && h_{i j}- \eta \partial_0 h_{i j}- \frac{1}{2} \eta^2 \partial_0^2 h_{i j}-\frac{1}{2} \eta^2\left(\partial_k \partial_i h_{j k}+\partial_k \partial_j h_{i k}-\partial_i \partial_j h-\partial_k^2 h_{i j}\right) \nonumber\\
 &+ & \delta_{ij} \left( - h + \eta\partial_0 h + \frac{1}{2} \eta^2 \partial_0^2 h -\frac{1}{2} \eta^2 \partial_l^2 h+\frac{1}{2} \eta^2 \partial_k \partial_l h_{k l}  \right) = \frac{T_{ij}}{H^2}\ .
\end{eqnarray}

\section{dilaton derivation}
\label{app,dilaton}
 In this section, we recover the vanishing dilaton in flat space without fixing the temporal gauge in the dictionary itself, but rather treating the physical solution as one that happens to have a vanishing $A_0$. The dilaton dictionary contains the contribution
\begin{eqnarray}
    A \circ A - \frac{1}{\square} \partial A \circ \partial A = \frac{g \pi }{4}\frac{\delta(k_0)}{k_0^2 }   + \frac{1}{k_0^2 - k^2} \frac{g^2 \pi }{4}\frac{\delta(k_0)}{k_0^2 }  k^2 = \frac{g^2 \pi }{4}\frac{\delta(k_0)}{k_0^2 -k^2  }  \ ,
\end{eqnarray}
where we utilize Eq.\eqref{eq,AicircAj}. 
Therefore, to get a vanishing dilaton, the ghost-antighost term should behave as 
\begin{eqnarray}
   \xi \partial_0 c \circ \bar c = \frac{\mathfrak{b}_1}{2 \mathfrak{b}_2} \frac{g \pi }{4}\frac{\delta(k_0)}{k_0^2 -k^2  }  \ . 
\end{eqnarray}
Notice under the inverse Fourier transformation, $\delta(k_0) $ picks up the $k_0 =0$ phase, and the above expression is equivalent to 
\begin{eqnarray}
    \xi \partial_0 c \circ \bar c = \frac{\mathfrak{b}_1}{2 \mathfrak{b}_2} \frac{g \pi }{4}\frac{\delta(k_0)}{ -k^2  } \ ,  
\end{eqnarray}
which will behave like $1/r$ in coordinate space. Plugging this ghost-antighost term back in the graviton's dictionary, we have
\begin{eqnarray}
\mathfrak{a}_1  A_i \circ A_j   + 2 \mathfrak{a}_{123} \left(\frac{\partial_i \partial j}{\partial_0^2 } - \delta_{ij } \right) = \left(\mathfrak{a}_1  - \frac{\mathfrak{b}_1}{\mathfrak{b}_2}\mathfrak{a}_{123}   \right) \frac{g \pi}{4} \frac{\delta(k_0)}{k_0^2} \frac{k_i k_j}{k^2} + \frac{\mathfrak{b}_1}{\mathfrak{b}_2}\mathfrak{a}_{123}    \frac{g \pi}{4} \frac{\delta_{ij}\delta(k_0)}{k^2} \ .
\end{eqnarray}
To map with the Fourier transformation of the graviton 
\begin{eqnarray}
    \mathcal{F}(h_{ij}) = r_s 8\pi^2 \frac{\delta(k_0)}{k_0^2} \frac{k_i k_j}{k^2} + r_s 8\pi^2 \frac{\delta_{ij }\delta(k_0)}{k^2} \ ,
\end{eqnarray}
obtained from Eq.\eqref{eq,gravitonaxial}, we have
\begin{eqnarray}
   \frac{g \pi}{4} \frac{\mathfrak{b}_1}{\mathfrak{b}_2}\mathfrak{a}_{123}  = 8\pi^2 r_s\ , \quad   \frac{g \pi}{4} \mathfrak{a}_{1} = 16 \pi^2 r_s\ . 
\end{eqnarray}
This gives an example of a convolution double copy through a general dictionary. 

\bibliography{ref}

\end{document}